\documentclass[prd,nofootinbib,twocolumn,showpacs]{revtex4}
\usepackage{graphics}
\usepackage{bm}

\begin{document}

\title{On the Detection of Magnetic Helicity}

\author{
Tina Kahniashvili$^{1,2}$ and Tanmay Vachaspati$^{3}$
}
\affiliation{
$^{1}$Department of Physics, Kansas State University, Manhattan, KS, USA.\\
$^{2}$Center for Plasma Astrophysics, Abasturnani Astrophysical Observatory,
Tbilsi, Georgia.\\
$^{3}$CERCA, Department of Physics, Case Western Reserve University,
10900 Euclid Avenue, Cleveland, OH 44106-7079, USA.}

\begin{abstract}
\noindent
Magnetic fields in various astrophysical settings may be helical and,
in the cosmological context, may provide a measure of primordial
CP violation during baryogenesis. Yet it is difficult, even in principle,
to devise a scheme by which magnetic helicity may be detected, except in
some very special systems. We propose that charged cosmic rays originating
from known sources may be useful for this purpose. We show that the
correlator of the arrival momenta of the cosmic rays is sensitive
to the helicity of an intervening magnetic field. If the sources
themselves are not known, the method may still be useful provided
we have some knowledge of their spatial distribution.
\end{abstract}

\maketitle


Magnetic fields pervade all astrophysical 
objects \cite{ZelRuzSok,SemSok05} and 
there are good theoretical reasons to believe that a weak magnetic field 
is present throughout the universe.
In astrophysical systems the magnetic field is often helical which means 
that the field lines are twisted (like corkscrews), or that closed magnetic
lines are linked. Mathematically, the average helicity density in a
volume $V$ is defined as:
\begin{eqnarray}
H = \frac{1}{V} \int_V d^3 x ~ {\bf A}\cdot {\bf B}.  
\nonumber
\end{eqnarray}
In cosmology, a number of scenarios predict the creation of a primordial
field with non-zero helicity. In the scenario discussed in 
Refs.~\cite{Vachaspati:1991nm,Vachaspati:1994xc,Cornwall:1997ms,
Vachaspati:2001nb}, a magnetic field is produced at the 
electroweak phase transition. The helicity of the magnetic field is 
related to the cosmological baryon asymmetry arising from CP violation 
in the fundamental particle physics theory, and the sign of the helicity 
is predicted to be left-handed \cite{Vachaspati:2001nb}. 
There are also several other scenarios for the generation of primordial 
helical magnetic fields that do not depend on the dynamics through 
a phase transition \cite{generation-helicity1,2,3,4,5,6,7}.


The helicity of magnetic fields in astrophysical jets can be deduced 
from the polarization of synchrotron radiation \cite{ensslin,valee}. 
In such situations, the velocity of electrons in the jets is known 
and this additional information is crucial to the determination of
helicity. In other situations, it is much harder to find the
helicity. For example, Faraday rotation only provides an estimate
of the line of sight component of the magnetic field. 
Even by observing the Faraday rotation from different sources, the
information is insufficient to estimate the helicity
\cite{ensslin03,campanelli,Kosowsky:2004zh}. 
An estimate of the helicity necessarily requires sensitivity to
all components of the magnetic field and hence it is a challenging
theoretical problem to devise means by which it may be 
measured. In Ref. \cite{pogosian02,cdk04,kr05} the imprint of cosmological 
magnetic helicity in parity-odd cross correlations of the cosmic 
microwave background (CMB) fluctuations was investigated, while in 
Ref. \cite{kgr05} it was shown that helicity would introduce
circular polarization of induced relic gravitational waves.
Both these potential signatures of helicity are limited to 
cosmological magnetic fields since they rely on properties of
the cosmic microwave background or on the cosmic gravitational 
wave background. Further, the signals are small for several reasons:
the cosmic magnetic field is constrained to be weaker
than $\sim 1$ nG, the CMB is polarized only at the 10\% level,
the tensor modes that enter parity-odd correlations are tiny, and
the gravitational waves are extremely weak.

In the present paper, we show that correlators of the arrival
momenta of charged cosmic rays from {\it known} sources carry
information about the helicity of the magnetic field through which
the charges propagate. The scheme has the advantage that it is not
restricted to cosmological magnetic fields, and it utilizes cosmic
rays which are abundant and well-studied \cite{dolag}.  
 The difficulty with our
scheme is that we do not normally know the source from which an
observed cosmic ray emanated. However, the scheme may be
extended to situations where we have some knowledge of the
distribution of sources {\it e.g.} if the sources are located
within a certain region of the disk of the galaxy. We have not yet 
explored this possibility in detail. For the present paper, we focus 
on establishing an observable that is sensitive to magnetic helicity. 
Further work is needed to decide if the observable that we propose 
is practically useful.

\begin{figure}
\scalebox{0.50}{\includegraphics{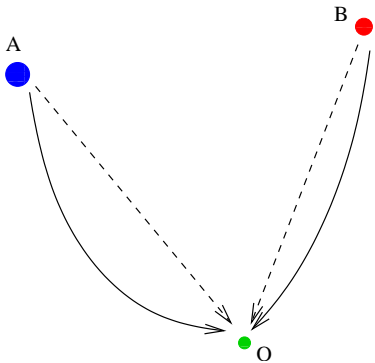}}
\caption{Two sources $A$ and $B$ emit charged particles that
are observed on Earth at $O$. If there was no ambient magnetic 
field, the particles would follow straight trajectories (dashed
lines). In the presence of a weak magnetic field, the trajectories
get bent (solid curves).}
\label{ABO}
\end{figure}

Consider a situation where there are two known sources
(A and B) that are emitting charged particles that arrive on
Earth. The particles would propagate along straight lines
from the sources to the Earth if there were no magnetic
field. However, the trajectories get bent by the weak magnetic 
field. We work to lowest order in the magnetic field
strength and consider the momenta
of the particles as being perturbed due to the magnetic field:
\begin{eqnarray}
{\bf P}_A = {\bf P}_{0A} + {\bf p}_A \ , \ \  
{\bf P}_B = {\bf P}_{0B} + {\bf p}_B 
\nonumber
\end{eqnarray}
where the $0$ in the subscript denotes an unperturbed momentum, 
and ${\bf p}_{A,B}$ are the momentum perturbations. The unperturbed
momenta are directed along the lines of sight to the sources and 
the magnitudes are completely determined by the energies 
of the charged particles. 

We are interested in the observable
\begin{equation}
\langle {P}_A^i(t_f) {P}_B^{i'}(t_f') \rangle 
=  {P}_{0A}^i {P}_{0B}^{i'} + \langle {p}_A^i(t_f) {p}_B^{i'}(t_f') \rangle  
\label{vcorrel}
\end{equation}
where $i,i'=1,2,3$ are spatial indices and $t_f$ and $t_f'$ denote arrival 
times from the two sources. 
The ensemble average refers to an average over 
many realizations of the magnetic field for the same locations of the 
two sources. We will discuss ways in which an ensemble average can be 
implemented toward the end of the paper. In writing Eq.~(\ref{vcorrel}) 
we have implicitly considered particles for which the energies are fixed.
Otherwise we would also have to average the unperturbed momenta since 
these depend on the energies of the particles. We have also taken 
$\langle {\bf p}_{A,B} \rangle =0$ which holds for a stochastic magnetic 
field with zero mean. 

The first term in Eq.~(\ref{vcorrel}) contains the unperturbed 
momenta. To evaluate this contribution, it is essential that 
we have some knowledge of the locations of the two sources.
We begin by assuming that we know the locations exactly and
later comment on the case where the sources are distributed in a
plane.  We now evaluate the correlator for the momentum perturbations 
\begin{eqnarray}
C^{ii'} = C^{ii'}({\bf X}_A, {\bf X}_B)
      \equiv \langle {p}_A^i (t_f) {p}_B^{i'} (t_f')\rangle .
\nonumber
\end{eqnarray}
where ${\bf X}_A$ and ${\bf X}_B$ are the vectors from $O$ to
sources $A$ and $B$ respectively.
We find that certain components of $C^{ii'}$ are sensitive only 
to the helicity of the intervening magnetic field and vanish
if the helicity is zero.

To proceed, we use the Lorentz force law to find ${\bf p}$ to
linear order in the magnetic field ${\bf B}$,
\begin{eqnarray}
{\bf p}(t) = {\bf p}_i  +
 q \int_{t_i}^t dt' ~
      {\bf v}_0 \times {\bf B}({\bf x}(t')),
\label{momentum}
\end{eqnarray}
where we have temporarily suppressed the subscripts specifying the
source for convenience. The unperturbed velocity 
${\bf v}_0$ may be related to the unperturbed momentum using 
${\bf v}_0 = {\bf P}_0/E$, where $E$ is the energy.
The initial momentum perturbation
${\bf p}_i \equiv {\bf p}(t_i)$ need not vanish and also
depends on the
intervening magnetic field. To determine ${\bf p}_i$
we integrate ${\bf P}={\bf P}_0 + {\bf p}(t)$ over proper time
\begin{eqnarray}
{\bf x} (t) &=& {\bf x}_i + \frac{{\bf P}_0 + {\bf p}_i}{E} (t-t_i)
              \nonumber \\
 &+& \frac{q}{E} \int_{t_i}^t dt' \int_{t_i}^{t'} dt'' ~
      {\bf v}_0 \times {\bf B}({\bf x}(t'')),
\label{position}
\end{eqnarray}
where ${\bf x}_i \equiv {\bf x}(t_i)$ and $E$ is the energy of
the particle.

We know that the charged particle arrives at the detector at
$t=t_f$. Taking the origin of the coordinate system to be at the
position of the detector, we find
\begin{eqnarray}
{\bf p}_i
  &=& 
    - \frac{q}{T} \int_{t_i}^{t_f} dt'
      \int_{t_i}^{t'} dt'' ~ {\bf v}_0 \times {\bf B}({\bf x}(t''))
\label{initialmomentum}
\end{eqnarray}
where $T \equiv t_f-t_i$ and we have used 
${\bf P}_0 = -E {\bf x}(t_i)/T$.
Going back to Eq.~(\ref{momentum}), we can write
\begin{equation}
{\bf p}(t_f) = 
      {\cal I} [{\bf v}_0 \times {\bf B}({\bf x})]
\label{momentumI}
\end{equation}
where the action of the operator ${\cal I}$ on a function $F(t)$
is defined by
\begin{eqnarray}
{\cal I} [F] &\equiv& - \frac{q}{T} \int_{t_i}^{t_f} dt'
                              \int_{t_i}^{t'} dt'' F(t'')
       + q \int_{t_i}^{t_f} dt' F(t') \nonumber \\
 &=&
 - \frac{q}{T} \int_{t_i}^{t_f} dt' [ 1 - T\delta (t'-t_f) ]
                              \int_{t_i}^{t'} dt'' F(t'')
\end{eqnarray}
Therefore
\begin{equation}
C^{ii'}  = {\cal I}_A {\cal I}_B
 \biggl [ \epsilon_{ijk} \epsilon_{i'j'k'} v_{0A}^j v_{0B}^{j'}
       \langle B^k ({\bf x}_A (t)) B^{k'} ({\bf x}_B (t')) \rangle
 \biggr ]
\label{Ccorrel}
\end{equation}
where, for generality, we consider different species of charged
particles arriving from sources $A$ and $B$ with different energies
$E_A$, $E_B$ and travel times $T_A$, $T_B$.


The auto-correlator of an isotropic, stochastic, time-independent
magnetic field can be written as \cite{my75} 
\begin{eqnarray}
\langle B_i({\bf {x + r}}) B_j({\bf {x}}) \rangle
&=& M_N (r) \left [ \delta_{ij} -  \frac{r_i r_j}{r^2} \right ]  +
M_L (r) \frac{r_i r_j}{r^2}
\nonumber\\ &+&
M_H(r) \varepsilon_{ijl} r_l,
\label{Bcorrel}
\end{eqnarray}
where $M_N(r)$, $M_L(r)$, and $M_H(r)$ are the correlation functions for 
the ``Normal'', ``Longitudinal'',   and ``Helical'' parts of the magnetic 
field. We assume that spacetime curvature can be neglected on the length
scales of interest and use a Minkowski metric, so $B_i =-B^i$. 
All correlation functions depend only on $r=|{\bf r}|$, reflecting the 
statistical isotropy of the field. The divergence-less condition requires 
\begin{eqnarray}
M_N(r) = \frac{1}{2r} \frac{d}{dr} (r^2 M_L(r)). 
\nonumber
\end{eqnarray} 
The ensemble averaging in Eq.~(\ref{Bcorrel}) is over 
all locations ${\bf x}$ but for fixed ${\bf r}$.

The magnetic field two-point correlation function is often given in 
Fourier space, so it is useful to express $M_N(r)$, $M_L(r)$, and $M_H(r)$ 
in terms of a magnetic field wave number ${\bf k}$ space power spectrum 
defined by:
\begin{eqnarray}
\langle B_i^\star({\bf {k}}) B_j({\bf {k^\prime}}) \rangle &=&
(2\pi)^3 \delta^{(3)}({\bf k} - {\bf{k^\prime}}) \nonumber \\
&\times&
\left[ P_{ij} {F}_N (k) +i \varepsilon_{ijl} \frac{k_l}{k} F_H(k)\right]
\nonumber
\end{eqnarray}
where the projector 
$P_{ij}({\mathbf{\hat k}})\equiv\delta_{ij}-\hat{k}_i\hat{k}_j$, 
and the unit vector $\hat{k}_i=k_i/k$.  $F_N(k)$ and $F_H(k)$ are the 
symmetric and helical parts of the magnetic field power spectrum, 
related to the average energy density and helicity of the magnetic 
field. The functions $F_N (k)$ and $F_H (k)$ can be related to the
correlation functions $M_N (r)$, $M_L (r)$ and $M_H (r)$ as
in Ref.~\cite{my75}.

The correlator $C^{ii'}$, Eq.~(\ref{Ccorrel}), may be decomposed 
into normal,  longitudinal, and helical parts, 
\begin{eqnarray}
C^{ii'} = C^{ii'}_N   + C^{ii'}_L + C^{ii'}_H   
\nonumber
\end{eqnarray}
The remaining calculation involves inserting Eq.~(\ref{Bcorrel})
into (\ref{Ccorrel}) and simplifying. Let us define
\begin{eqnarray}
{\bf n} = {\bf v}_{0A} \times {\bf v}_{0B}. 
\label{normal}
\end{eqnarray}
A straightforward computation gives the correlator induced by the 
normal component of the magnetic field power spectrum, 
\begin{eqnarray}
C^{ii'}_N  &=& 
  {\cal I}_A {\cal I}_B \biggl [ M_N (r) \biggl (  
    \delta^{ii'} {\bf v}_{0A}\cdot {\bf v}_{0B} 
        - {\bf v}_{0A}^{i'} {\bf v}_{0B}^i \nonumber \\
&&\hskip 2.0cm - [{\bf v}_{0A} \times {\bf {\hat r}}]^i 
                 [{\bf v}_{0B} \times {\bf {\hat r}}]^{i'}
                                         \biggr ) \biggr ]
\end{eqnarray}
where ${\bf r}(t, t') = {\bf x}_A (t) - {\bf x}_B (t')$ and the unit 
vector ${\bf {\hat r}} = {\bf r}/r$. 
The longitudinal piece of correlator is 
\begin{eqnarray}
C^{ii'}_L = 
  {\cal I}_A {\cal I}_B \biggl [ M_L (r) 
[{\bf v}_{0A} \times {\bf {\hat r}}]^i [{\bf v}_{0B} \times {\bf{\hat r}}]^{i'}
                                         \biggr ]
\end{eqnarray}
Similarly for the helical component we get:
\begin{eqnarray}
C^{ii'}_H  &=& 
  {\cal I}_A {\cal I}_B \biggl [ M_H (r) \biggl (  \epsilon^{ii'l} 
         [ ({\bf v}_{0A}\cdot {\bf r}) v_{0B}^l + 
             v_{0A}^l ({\bf v}_{0B} \cdot {\bf r}) ] \nonumber \\
&&\hskip 2.5cm + [ {r}^i {n}^{i'}+ {r}^{i'} {n}^i ] \biggr ) \biggr ]
\nonumber
\end{eqnarray}

The helical part of correlator, $C_H^{ii'}$, vanishes for $i=i'$, and
the trace of the momentum perturbation correlator contains contributions 
only from the normal and longitudinal parts of the magnetic field spectrum: 
\begin{eqnarray}
C^{{\rm{TR}}} &=& 
  {\cal I}_A {\cal I}_B \biggl [ 2 M_N (r) {\bf v}_{0A}\cdot {\bf v}_{0B} 
                    \nonumber\\ 
&& \hskip -1.0cm + (M_L(r) - M_N(r)) 
 [({\bf v}_{0A}\cdot {\bf v}_{0B})- ({\bf v}_{0A}\cdot {\bf {\hat r}})
({\bf v}_{0B}\cdot {\bf {\hat r}}) ] 
         \biggr ]
\nonumber
\label{trace}
\end{eqnarray}
Let us take our coordinate system so that the triangle $ABO$ lies
in the $xy-$plane (see Fig.~\ref{ABO}) and ${\bf n}$ is in the 
${z}-$direction. We find that all components of the helical
correlator vanish except for:
\begin{eqnarray}
C^{iz}_H  &=& 
- 2 ({\bf v}_{0B} \cdot {\bf M}_H)
                        ({\bf v}_{0A} \times {\bf {\hat z}})^i
\label{Czi} \\
C^{zi}_H  &=& 
+ 2 ({\bf v}_{0A} \cdot {\bf M}_H)
                        ({\bf v}_{0B} \times {\bf {\hat z}})^i
\label{Ciz}
\end{eqnarray}
where 
\begin{equation}
{\bf M}_H \equiv {\cal I}_A {\cal I}_B [M_H (r(t,t')) {\bf r}(t,t')]
\label{vecMHdefn}
\end{equation}
After doing the integrations, any dependence of ${\bf M}_H$ 
on $t_f$ and $t_f'$ can be traded for a dependence on the position of the 
particles at the final time using: 
${\bf x}_A (t_f) = {\bf x}_B (t_f') = 0$. A neat combination of the
two components in Eqs.~(\ref{Czi}) and (\ref{Ciz}) is
\begin{equation}
{\bf C}_H^i \equiv \frac{1}{2} [ C_H^{iz} + C_H^{zi} ]
          =  {\bf M}_H^i | {\bf v}_{0A} \times {\bf v}_{0B} |
\end{equation}
where we have used ${\bf M}_H \cdot {\hat {\bf z}} =0$.

The normal component $C^{ii'}_N$ vanishes  
if $i$ or $i'$ (but not both) are in the 
${z}-$direction. The trace of momentum correlator 
has the contribution only from $M_N$ and $M_L$, and in $xy-$plane 
$C^{\rm{Tr}}_{{\rm{xy}}}=\sum_{\alpha}C^{\alpha \alpha}$ ($\alpha=x, y$)
depends only on the normal component $M_N(r)$, 
\begin{eqnarray}
C^{\rm{Tr}}_{{\rm{xy}}} = 
  {\cal I}_A {\cal I}_B [ M_N (r) ] ~ {\bf v}_{0A}\cdot {\bf v}_{0B}
\nonumber
\end{eqnarray}
The longitudinal component $C^{ii'}_L$
vanishes if $i\neq i'$ and has only one non-zero component when $i=i'$ 
is along the $z$ direction, 
\begin{eqnarray}
C^{zz}_L  = 
  {\cal I}_A {\cal I}_B \biggl [ M_L (r) \{  {\bf v}_{0A}\cdot {\bf v}_{0B}
 - ({\bf v}_{0A} \cdot {\bf {\hat r}})({\bf v}_{0A} \cdot {\bf {\hat r}})
                                         \} \biggr ]
\nonumber
\end{eqnarray}
On the other hand, $C^{ii'}_H$ is non-zero only if
one (and only one) of $i$, $i'$ is along the $z-$direction.
This can be understood on physical grounds as follows. If the magnetic
field is not helical, a charged particle is as likely to be deflected
in the $+z$ direction as it is to be in
the $-z$ direction by the stochastic magnetic field. By symmetry,
the components $C^{iz}$ must then vanish. However, a helical magnetic
field breaks the symmetry and these components become non-zero.
So we see that only the helical contribution enters the $xz$, $yz$, 
$zx$ and $zy$ components of $C^{ii'}$, and further, only the 
non-helical contributions enter the other components. Therefore the 
normal and helical pieces of the correlator do not mix.

The other components of the correlator ({\it e.g.} $C_{xy}^{\rm{Tr}}$)
can be used to find the normal and longitudinal correlation functions,
$M_N$ and $M_L$. Correlations of the rotation measure due to Faraday 
rotation of polarized light from different sources can also be used 
as an independent method to determine $M_N$ and $M_L$ \cite{ensslin03}.


The above analysis relies on an average over an ensemble of
random magnetic field realizations for fixed source and 
detector positions (triangle ABO in Fig.~\ref{ABO}). In our 
universe, however, only one realization of the magnetic field
is available and so we need to discuss a practical scheme for 
doing the ensemble average and thus estimating the correlator,
$C^{ii'}$. The ensemble average would necessarily involve
averaging over many pairs of sources that we denote by
$(A,B)_\alpha$, $\alpha = 1,2,...,N$ from which we assume
that cosmic rays are observed with average momenta 
$({\bf P},{\bf P}')_\alpha$. Going back to the magnetic field 
correlation function, Eq.~(\ref{Bcorrel}), we see that the
ensemble average is over all ${\bf x}$ for fixed point
separation ${\bf r}$. Since $C^{ii'}$ depends on integrals of 
the magnetic field along the line of sight, this suggests
that we find pairs of sources such that ${\bf r}(t,t')$
is the same function of $t$ and $t'$ for all of them. Together 
with the constraint that the position of the detector is fixed 
at O, such an ensemble has only one element and is not useful. 
Instead of holding ${\bf r}(t,t')$ fixed, it is more useful
to choose pairs with a less restrictive condition. There are
many such possible conditions, each with its own advantage. 
As an example, one such condition is that the source separation 
vector, ${\bf \Delta}\equiv {\bf X}_B-{\bf X}_A$, be held 
fixed for all pairs in the ensemble. The corresponding 
observable is
\begin{equation}
{\cal P}^{ii'}({\bf \Delta}) = \frac{1}{N} \sum_{\alpha=1}^N 
                       {\bf P}_\alpha^i {\bf P}_\alpha'^{i'}
\label{estimator}
\end{equation}
This observable quantity is an estimator for an average of the 
total correlator $C^{ii'}$ discussed above, where the 
average is over sources with fixed separation vector $\bf \Delta$.
As we have noted before, the helical part, and only the
helical part, of the magnetic
field enters certain components of the total correlator
(see Eqs.~(\ref{Czi}) and (\ref{Ciz})). So to get an estimator
for the helical components of the correlator we should only
look at those components of the estimator (\ref{estimator})
that involve momentum in a direction in the source-observer 
plane and the second momentum in a direction perpendicular to 
this plane. Since different pairs of sources (with the same 
${\bf \Delta}$) may lie in different planes, an estimator for the 
helical part of the correlator is
\begin{equation}
{\cal P}_H ({\bf \Delta}) = \frac{1}{N} \sum_{\alpha=1}^N 
          ({\bf n}_\alpha \cdot {\bf P}_\alpha )
          ({\bf e}_\alpha \cdot {\bf P}_\alpha')
\label{helicalestimator}
\end{equation}
where ${\bf n}_\alpha$ is the normal to the plane of the source pair 
labeled by $\alpha$ and the observer as defined in Eq.~(\ref{normal}), 
and ${\bf e}_\alpha$ is some chosen unit vector within the plane.

To extract the magnetic helicity from a measurement of 
${\cal P}_H ({\bf \Delta})$, we must evaluate the corresponding
theoretical quantity, which is given by
\begin{eqnarray}
{\bf \bar{C}}_{H} ({\bf \Delta})= 
\frac{1}{V} \int_V d^3 X_A {\bf C}_{H} ({\bf X}_A,{\bf X}_A + 
{\bf \Delta}) 
\nonumber
\end{eqnarray} 
The above integral is quite involved but can be done numerically 
for different choices of ${\bf M}_H (r)$. 

To summarize, a non-vanishing observed value of 
${\cal P}_H ({\bf \Delta})$
will give a measure of ${\bf \bar{C}}_{H} ({\bf \Delta})$ 
and hence will lead to the magnetic helicity, ${\bf M}_H (r)$.

 
At present we do not have any known sources of charged cosmic 
rays. However, it is likely that a large fraction of cosmic
rays that we see arise in the galactic disk (or sources 
confined in the cosmic large-scale structure \cite{ds05}). It may be possible 
to usefully extend the ensemble average to include pairs of 
locations in the galactic disk. Such an averaging could 
still yield information about the helicity of the galactic magnetic 
field. We plan to consider this extension of our result in future 
work, together with other observational issues.

\begin{acknowledgments} 
We thank Torsten Ensslin, Francesc Ferrer, and Bharat Ratra for 
helpful discussions and suggestions, and Corbin Covault, 
Dario Grasso, and Arthur Kosowsky for comments. This work was 
supported by the U.S. Department of Energy and NASA at 
Case Western Reserve University. T.K.  acknowledges support from 
DOE EPSCoR grant DE-FG02-00ER45824.
\end{acknowledgments}

\end{document}